\documentclass[preprint2]{aastex}
\usepackage{amsmath}
\usepackage{amssymb}
\usepackage{graphicx}
\usepackage{array}
\usepackage{lscape}

\providecommand{\tabularnewline}{\\}

\def\aj{Astron.\ J.}

\def\apj{Astrophys.\ J.}
\def\apjl{Astrophys.\ J.\, Lett.}
\def\apjs{Astrophys.\ J.\, Suppl.\ Ser.}

\def\mnras{Mon.\ Not.\ R.\ Astron.\ Soc.}

\def\nat{Nature}

\def\prd{Phys.\ Rev.\ D}

\begin{document}

\title{Detecting the Rise and Fall of the First Stars by Their Impact on Cosmic Reionization}

\author{Kyungjin Ahn\altaffilmark{1}}
%%\affil{Department of Earth Sciences, Chosun University, Gwangju 501-759, Korea}
%%\email{kjahn@chosun.ac.kr}

\author{Ilian T. Iliev\altaffilmark{2}}
%%\affil{Astronomy Centre, Department of Physics \& Astronomy, Pevensey II Building, University of Sussex, Falmer, Brighton BN1 9QH, United Kingdom}

\author{Paul R. Shapiro\altaffilmark{3}}
%%\affil{Department of Astronomy and the Texas Cosmology Center, University of Texas, Austin, TX 78712-1083, U.S.A}

\author{Garrelt Mellema\altaffilmark{4}}
%%\affil{Department of Astronomy \& Oskar Klein Centre, Stockholm University, Albanova, SE-10691 Stockholm, Sweden}

\author{Jun Koda\altaffilmark{5}}
%%\affil{Centre for Astrophysics \& Supercomputing, Swinburne University of Technology, Hawthorn, Victoria 3122, Australia}

\and

\author{Yi Mao\altaffilmark{3}}

\altaffiltext{1}{Department of Earth Sciences, Chosun University,
Gwangju 501-759, Korea; kjahn@chosun.ac.kr}
\altaffiltext{2}{Astronomy Centre, Department of Physics \& Astronomy, 
Pevensey II Building, University of Sussex, Falmer, Brighton BN1 9QH, 
United Kingdom}
\altaffiltext{3}{Department of Astronomy, University of Texas, Austin, TX 78712-1083, U.S.A}
\altaffiltext{4}{Department of Astronomy \& Oskar Klein Centre, Stockholm University, Albanova, SE-10691 Stockholm, Sweden}
\altaffiltext{5}{Centre for Astrophysics \& Supercomputing, Swinburne
University of Technology, Hawthorn, Victoria 3122, Australia}

\begin{abstract}

The intergalactic medium was reionized before redshift $z \sim 6$,
most likely by starlight which escaped from early galaxies.  
The very first stars formed when hydrogen molecules (H$_2$) cooled gas inside
the smallest galaxies, \emph{minihalos} of mass between $10^5$ and
$10^8 \,M_\odot$. Although the very first stars began
forming  inside these minihalos before redshift $z \sim 40$,
their contribution has, to
date, been ignored in large-scale simulations of this cosmic reionization.
Here we report results
from the first reionization simulations to include these first stars and
the radiative feedback that limited their formation, in a volume large
enough to follow the crucial spatial variations that influenced the process
and its observability.  We show that,
while minihalo stars stopped far short of fully ionizing
the universe, reionization began much earlier
\emph{with} minihalo sources than
without, and was greatly extended, which boosts the intergalactic
electron-scattering optical depth and the 
large-angle polarization fluctuations of the cosmic microwave background
significantly.  Although within current \emph{WMAP} uncertainties, this
boost should be readily detectable by \emph{Planck}.  If reionization
ended as late as $z_{\rm ov} \lesssim 7$, as suggested by other observations,
\emph{Planck} will thereby see the signature of the first stars at
high redshift, currently undetectable by other probes. 

\end{abstract}


%%% Paul's revision: the section before Section 1 rewritten 

\section{Introduction}
\label{sec:intro}
The theory of reionization has not yet advanced to the point
of establishing unambiguously its timing and the relative contributions
to it from galaxies of different masses.  
In a Cold Dark Matter (``CDM'') universe, these early galactic sources
can be categorized by their host
halo mass into minihalos (``MHs'') and
``atomic-cooling'' halos (``ACHs''). MHs have masses
$M \sim 10^5 - 10^8$ $M_\odot$ and virial
temperatures $T_{\rm vir}\sim 10^4$ K, and thus
molecular hydrogen (H$_2$) was necessary to cool the gas below this
virial temperature to begin star formation. ACHs have 
$M\gtrsim 10^8$ $M_\odot$ and $T_{\rm vir}\gtrsim 10^4$ K, for which 
H-atom radiative line cooling alone was sufficient to support star formation.  
The ACHs can be split further into low-mass 
atomic-cooling halos (``LMACHs''; $M\sim 10^8 - 10^9$ $M_\odot$), for which the gas pressure of the 
photoionization-heated intergalactic medium (``IGM'') in an ionized patch 
prevented the halo from capturing the gas it needed to form stars, and 
high-mass atomic-cooling halos (``HMACHs''; $M \gtrsim 10^9 M_\odot$),
for which gravity was strong enough to overcome   
this ``Jeans-mass filter'' and form stars even in the ionized patches.  

Once starlight escaped from galactic halos into the 
IGM to reionize it, the ionized patches (``H II regions'') of the IGM became
places in which star formation was suppressed in both MHs and LMACHs.
At the same time, UV starlight at energies in the range 11.2 -- 13.6 eV
also escaped from the halos, capable of destroying the H$_2$ molecules 
inside MHs through Lyman-Werner band (``LW'') dissociation, even in the neutral
zones of the IGM.  This dissociation eventually prevented further
star formation in some of the MHs where the background intensity was
high enough. Early estimates, in fact, suggested that this would have made the MH 
contribution to reionization small \citep{Haiman2000}, and, until
now, large-scale simulations of reionization have neglected them
altogether.

In this letter we report the first radiative transfer (RT) simulations
of reionization to include \emph{all three} of the mass categories of 
reionization source halos, along with their radiative suppression,
in a simulation volume large enough to capture both the global mean 
ionization history and the observable consequences of its evolving 
patchiness in a statistically meaningful way\footnote{First results of
these simulations were briefly summarized in \citet{Ahn2010a}.}. 
We overcame the limitation 
of previous large-volume simulations by applying a newly developed 
sub-grid treatment to include MH sources (section~\ref{sec:methods}), and 
calculating the transfer of LW-band radiation self-consistently with
the source population using the scheme of
~\citet{Ahn2009}.

\section{Methods}
\label{sec:methods}
We performed a cosmological N-body simulation of structure formation with 
$3072^{3}$ particles in a $114/h\,{\rm Mpc}$ simulation box, using
the WMAP5 background cosmology \citep{Dunkley2009}.
For this we used the code CubeP$^{3}$M
(\citealt{PMFAST,2008arXiv0806.2887I}; J. Harnois-Deraps et al. 2012, in preparation), 
in which the gravity is computed by a P$^{3}$M 
(particle-particle-particle-mesh) scheme.
The simulation was started at redshift $z=300$ and run to 
$z=6$. N-body data  were recorded at 86 equally-spaced 
times (every $11.53\,{\rm Myrs}$) from $z=50$ to $z=6$. Each data
time-slice was then used to create matter density  
fields by smoothing the particle data
adaptively onto a uniform 
mesh -- or an ``RT grid'' -- of $256^3$ cells.
All cosmological halos with $M\ge 10^{8}\,M_{\odot}$
(corresponding to 20 particles or more), and thus both LMACHs
and HMACHs, were identified on-the-fly 
using a spherical overdensity halo finder with overdensity of
$\Delta=178$ with respect to the mean.

\begin{figure*}[ht]
\begin{center}
\hspace*{-0.6cm}\resizebox{18cm}{!}{\includegraphics{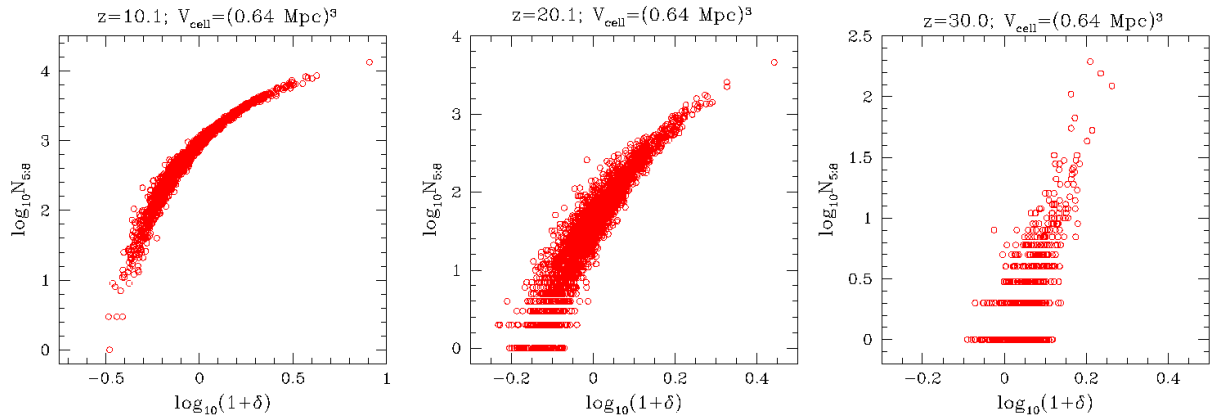}}
\end{center}
\caption{Correlation between the total number of MHs per RT cell 
($N_{5:8}$) and the cell density in units of the mean density 
($1+\delta$), based on a $6.3/h\,{\rm Mpc}$ box N-body simulation which
resolves all halos with $M\geq10^{5}\, M_{\odot}$. Plots are for
correlations at three different redshifts, $z=$30, 20.1 and 10.1 from
left to right. The volume
of the RT cell is $(0.64 {\rm Mpc})^{3}$.
\label{fig:delta-Nmini}}
\end{figure*}

Because MHs are too small to
be resolved in our simulation box, our RT grid was populated with
MHs through a newly-developed sub-grid model, as
follows. We started with a separate, high-resolution N-body
simulation of structure formation in a box with $(6.3/h~\rm
Mpc)^3$ volume and $1728^3$ particles, which resolved all MHs
with $M\ge 10^{5}M_{\odot}$ with 20 particles or more. We then
partitioned the box into a uniform grid of $14^{3}$ cells, such that
each cell is the same size as one of the RT grid cells
in our main, $114/h$~Mpc simulation box, and calculated cell density
and the total number of MHs per cell.
A strong and tight correlation
between the number of MHs located in a cell and its density is
observed (Fig.~\ref{fig:delta-Nmini}). The best fit to this correlation at
each redshift was then used as the total number of MHs in each grid
cell of $114/h$~Mpc box.

Based on these structure formation results for the IGM density field
and the source dark matter halos, we then calculated the  
radiative transfer of H-ionizing and ${\rm H}_2$-dissociating photons (see
Table~\ref{table:param} for the RT simulation parameters). The
stars inside ACHs are assumed to produce $g_{\gamma}$ ionizing
photons per baryon every 10 Myrs, where 
$g_{\gamma}\equiv f_\gamma/(t_{\star}/10\,{\rm Myr})$, and where
$f_\gamma \equiv  f_{{\rm e}}f_{\star}N_{{\rm i}}$,  
$f_{{\rm e}}$ is the escape fraction of ionizing photons, $f_{\star}$
is the star formation efficiency,
and $N_{{\rm i}}$ is the number of 
ionizing photons per stellar baryon produced over the star's lifetime
$t_{\star}$ -- we use $t_{\star}=11.53\,{\rm Myr}$ for both HMACHs and
LMACHs, and $t_{\star}=1.92\,{\rm Myr}$ for MHs. We assign one Pop~III star per
MH, motivated by numerical simulations of first star  
formation inside MHs, which find that typically one Pop III star 
with a mass between 100 and $1000\, M_{\odot}$ forms per MH in the absence
of strong soft UV radiative feedback \citep{2002ApJ...564...23B,
2002Sci...295...93A,Yoshida2006}. At each cell, only those MHs which are \emph{newly}
collapsed every 1.92~Myrs are assumed to host Pop III
stars to roughly approximate the disruptive radiative and
mechanical feedback by the first star and its by-products (such as a supernova)
on the halo gas (for this, we create ``morphed''
density fields every 1.92~Myrs by linearly interpolating the N-body
density fields in time which are separated by a time interval of
11.53~Myrs, and finite-difference corresponding minihalo populations
on each cell).
Star formation in MHs is further suppressed when the local LW
background -- calculated at each time step in 3D using the scheme by
\citet{Ahn2009}, but now considering both ACHs and MHs and also improved in speed
using the fast Fourier transform (FFT) scheme --
reaches a certain threshold $J_{{\rm LW,\, th}}$. At present the precise
value of this threshold is not well determined, 
but the typical values found by high-resolution simulations of
MH star formation are $J_{{\rm LW,\, th}} 
=[0.01-0.1]\times 10^{-21}\,{\rm erg}\,{\rm s}^{-1}\,{\rm cm}^{-2}\,{\rm sr}^{-1}$
\citep{Machacek2001,yoshida2003,Oshea2008}. We adopted a constant value
chosen from this range for each simulation. Even though stars may still form by ${\rm H}_2$
  cooling, when $J_{\rm LW}>J_{{\rm LW,\, th}}$, in MHs in
mass range $M\simeq 2\times 10^{6} - 10^{8} \,M_{\odot}$ 
and $M\simeq 10^{7} - 10^{8} \,M_{\odot}$ 
at $J_{{\rm LW,\, th}} 
\simeq 0.01\times 10^{-21}\,{\rm erg}\,{\rm s}^{-1}\,{\rm
  cm}^{-2}\,{\rm sr}^{-1}$ and $J_{{\rm LW,\, th}} 
\simeq 0.1\times 10^{-21}\,{\rm erg}\,{\rm s}^{-1}\,{\rm
  cm}^{-2}\,{\rm sr}^{-1}$ respectively (e.g. \citealt{Oshea2008}), we neglect
their contribution because they constitute only a small fraction of
the whole MH population.

The simulations with ACHs only (and without LW radiative transfer) are described in
~\citet{selfregulated:new}. Simulation parameters are given in
Table~\ref{table:param}. 

\section{Role of the first stars during cosmic reionization}

\begin{figure*}[ht]
\includegraphics[width=\textwidth]{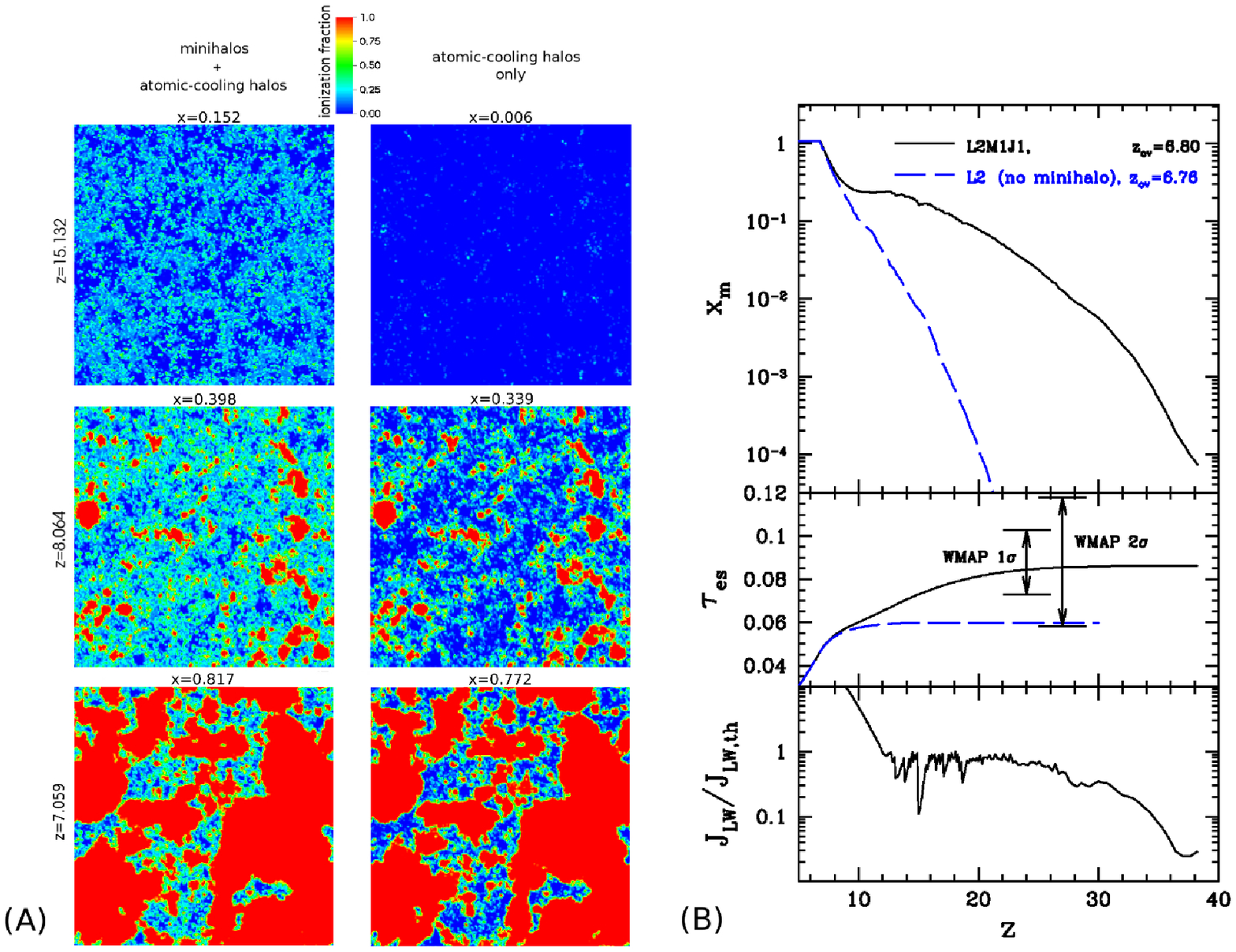}
\caption{(A) Maps of evolving hydrogen-ionized fractions at different
redshifts (rows), for our fiducial model with MH sources included, L2M1J1 (1st column), vs. the
corresponding reference model with only atomically-cooling halos,
case L2 (2nd column). The slices are $0.45/h$ Mpc-thick. Color represents
linearly-scaled ionized fraction from $0$ (blue) to $1$ (red). 
%%%%%% LW map violates ApJL figure limit. Removed. %%%%%
%(B) Maps of Lyman-Werner band intensity $J_{{\rm LW}}$ for case L2M1J1,
%before and after LW suppression takes place. Black
%contours indicate $J_{{\rm LW},\,{\rm th}}$ (which for case L2M1J1 is set at 
%$0.1\times10^{-21}\,{\rm erg}\,{\rm s}^{-1}\,{\rm cm}^{-2}\,{\rm
%sr^{-1}}$).
%%%%%%%%%%%%%%%%%%%%%%%%%%%%%%%%%%%%%%%%%%%%%%%%%%%%%%%%%%%%%%%%%
(B) (top) Globally-averaged history of the mass-weighted ionized fraction
for models L2M1J1 (black, solid) and L2 (blue, dashed). (middle)
$\tau_{\rm es}$ integrated from $z=0$ to redshift 
$z$ for L2 and L2M1J1. (bottom) Evolution of the mean 
$J_{{\rm LW}}$ in units of $J_{{\rm LW},\,{\rm th}}$
for Case L2M1J1. \label{fig:globalx}}
\end{figure*}

\begin{figure*}[ht]
\includegraphics[width=\textwidth]{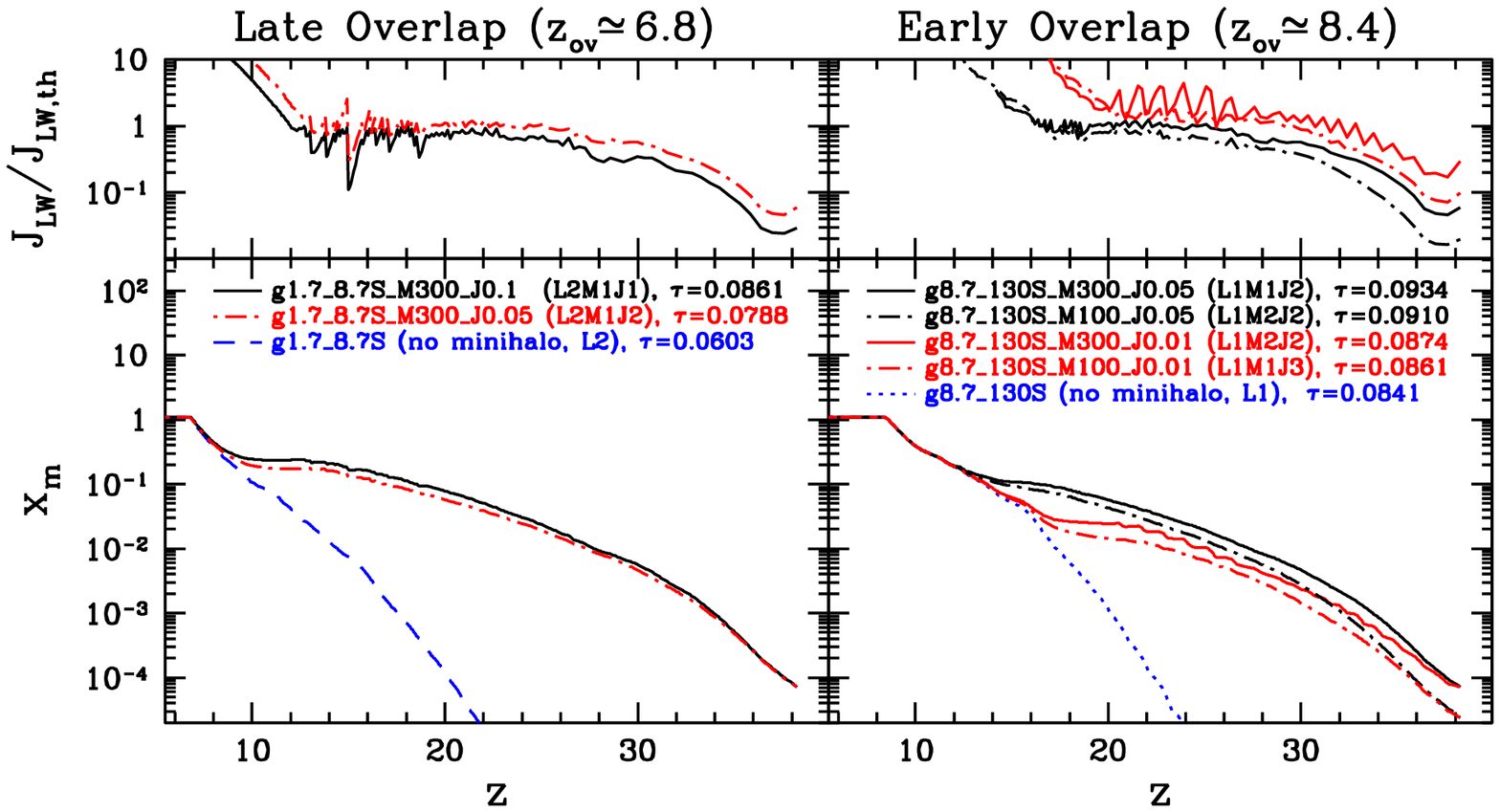}
\caption{Model dependency of the history of cosmic reionization.  In addition
  to cases L1 and L2, which do not account for  
MHs, we show predictions for MH-included cases by 
parametrizing the star formation inside MHs through $M_{*,\,{\rm III}}$ 
(mass of the Pop III star) and $J_{{\rm LW},\,{\rm th}}$ (threshold 
Lyman-Werner intensity). (Left) Late-overlap models. (Right) Early-overlap models.
\label{fig:models}}
\end{figure*}

We demonstrate the effects of the first stars by direct comparison
of the results from two simulations, a fiducial case which includes
all ionizing sources down to the first stars hosted by MHs (Case
L2M1J1) and a corresponding reference case which includes the larger,
atomically-cooling halos with exactly the same properties, but no
MH sources (Case L2, previously presented in 
~\citealt{selfregulated:new}). Our results show that
the early reionization history is completely dominated by the first
stars, while the late (redshift $z\lesssim 10$) history is driven by the stars 
inside HMACHs (Figs~\ref{fig:globalx}A and ~\ref{fig:globalx}B, top 
panel). The very first stars 
start to form inside MHs at redshift $z\simeq40$, and dominate the reionization 
process until $z\simeq10$ but through self-regulation which slows
their contribution to reionization\footnote{The oscillation of $J_{\rm
    LW}$ around the plateau at $J_{\rm LW} / J_{{\rm LW},\,{\rm th}}
  \sim 1$ observed in Figs.~\ref{fig:globalx}B and \ref{fig:models}  
is a numerical artifact which occurs because LW suppression locks the
MH star formation rate onto the level that keeps $J_{\rm LW} = J_{{\rm
    LW},\,{\rm th}}$, and the simulation time
step (1.92~Myrs) is comparable to the MH formation time scale.} (Fig.~\ref{fig:globalx}B).
Although the abundance of ACHs
rises exponentially, they 
remain relatively rare, and thus sub-dominant, until $z\simeq 10$. After
redshift $z\simeq 8$, though, the two reionization histories become largely 
indistinguishable, because the same HMACHs then dominate reionization
and push $J_{\rm LW}$ 
\emph{above} $J_{{\rm LW},\,{\rm th}}$  (at $z\simeq 12$) 
halting MH star formation altogether, long before the MHs can complete
reionization on their own.

Nonetheless, the MH sources (the first stars) can have quite 
a dramatic effect on the electron-scattering optical depth 
$\tau_{{\rm es}}$. 
While intergalactic 
H~II regions fully overlap (at redshift $z_{\rm ov}$, here defined as
when the mass-weighted mean ionized fraction in the IGM, $x_m$, first surpassed $99\%$) 
at almost identical redshifts 
$z_{{\rm ov}}\simeq 6.8$ with (L2M1J1) or without (L2) MHs, 
the early rise of $x_m$ with
MH sources boosts the optical depth by as much as $47\,\%$ relative to that without
MH sources: $\tau_{{\rm es}}=0.0861$ for L2M1J1, while
$\tau_{{\rm es}}=0.0603$ for L2. 
This satisfies the current observational constraints on reionization:
(1) reionization ended no earlier than redshift
$z=7$ ~\citep{2006AJ....132..117F,Ota2010,Mortlock2011,Bolton2011,Pentericci2011},
and (2) $\tau_{{\rm es}}=0.088\pm 0.015$ at $68\,\%$ confidence level \citep{wmap7_cosmo_parms}.
Predicted values of $\tau_{{\rm es}}$ and $z_{{\rm ov}}$ are model-dependent,
and thus we tested the robustness of our conclusions by varying the physical 
parameters of MHs and ACHs (Fig.~\ref{fig:models}).

The first stars, born inside MHs, imprint a
distinctive pattern on the global reionization history. 
For example,
in case L2M1J1, when the LW-plateau ended, reionization briefly stalled,
since MHs no longer formed the stars which replenished the ionizing
background and only ACH sources remained, thereafter; the ACH
contribution took a bit more time to climb enough to move
reionization forward again. This explains the brief ``$x_m$-plateau''
from $z\sim 12$ to $z\sim 10$ in Fig.~\ref{fig:globalx}B for case
L2M1J1, while in case L2, $x_m$ grows continuously without showing such
a plateau (Fig.~\ref{fig:globalx}B). This feature is generic (see Figs
~\ref{fig:models} and \ref{fig:ee}B for different sets of parameters
we explored).
Reionization histories \emph{without} MH sources, modelled
either by large-scale RT
simulations
\citep{2003MNRAS.343.1101C,2006MNRAS.369.1625I,Iliev2007,Trac2007,McQuinn2007,Zahn2007,selfregulated:new} 
or semi-analytical calculations \citep{Haiman2000,Zahn2007}, are all
similar in that respect. 

Reionization histories \emph{with} MH
sources calculated here, however, 
find an ionization plateau phase. 
Previous studies that considered MH stars and their impact were not
able to settle the issue of their global effect on reionization.
This is either because they simulated volumes much too small to represent a
fair portion of the Universe
\citep{2002ApJ...575...33R,Sokasian2004,Yoshida2006,Ricotti2008,Johnson2012}, or else treated
reionization by a semi-analytical, 1-zone, homogeneous approximation
(either \emph{with} LW suppression included 
\citep{Haiman2000,2005ApJ...634....1F} or \emph{without}
\citep{1994ApJ...427...25S,Wyithe2003,Haiman2006,Wyithe2007}, which
cannot capture its
innate spatially inhomogeneous nature, or made a semi-analytical
approximation that accounted statistically for spatial inhomogeneity
but without LW suppression \citep{2006ApJ...649..570K}.

We find that the global reionization history at $z\lesssim20$ depends on 
$J_{{\rm LW},\,{\rm th}}$ more strongly than $M_{\star,\,{\rm III}}$.
This is due to the very nature of self-regulation of the first star
formation. The larger the $J_{{\rm LW},\,{\rm th}}$ is, the weaker the 
suppression of star formation becomes, thereby temporarily hastening 
the progress of reionization. If $M_{\star,\,{\rm III}}$ is
smaller \citep{Turk2009,Stacy2010,Greif2011a,Stacy2012} than those simulated here, those
stars produce less ionizing 
and LW radiation and the  
resulting suppression is weaker, which partly compensates for the lower
emission per star. 
Similar type of compensation would occur also when the number of
MHs with a potential to form the first stars are smaller than our
estimate, due to the relative offset of baryonic gas from some of the
MHs \citep{Tseliakhovich2010,Greif2011}. 

\section{Probing the first stars with {\em Planck}}
We thus conclude that the first stars hosted by MHs likely made an important contribution
to reionization. But how can we probe them observationally? While
significant, the effects of the first stars are largely confined to
the early stages of reionization, at redshifts $z>10$, which puts
them beyond the reach of most current instruments. 
Recent observations by the South Pole Telescope (SPT) have been used to
place an upper limit on the kinetic Sunyaev-Zel'dovich (kSZ) effect from the
epoch of reionization \citep{Reichardt2011}.  While it has been suggested
that this restricts the duration of reionization \citep{Zahn2011}, we
will show elsewhere (H. Park et al. 2012, in prep) that the kSZ signal
from our fiducial case, L2M1J1, is well below the observed upper bound.
The combined effect
of the first stars will be reflected in the cosmic microwave 
background (CMB) polarization anisotropies at large scales. The
current best constraints on $\tau_{{\rm es}}$ by the \emph{WMAP} 
satellite $\tau_{{\rm es}}$ are still relatively weak, and thus 
models with low-$\tau_{{\rm es}}$ values like L2 are still acceptable 
at the $2\sigma$ (95\%) confidence level (Fig.~\ref{fig:globalx}B, 
middle panel). However, through the far more precise measurement of the CMB
polarization by the \emph{Planck} mission we should
be able to discern the influence of the first stars on 
reionization. 

\begin{figure*}[ht]
\includegraphics[width=\textwidth]{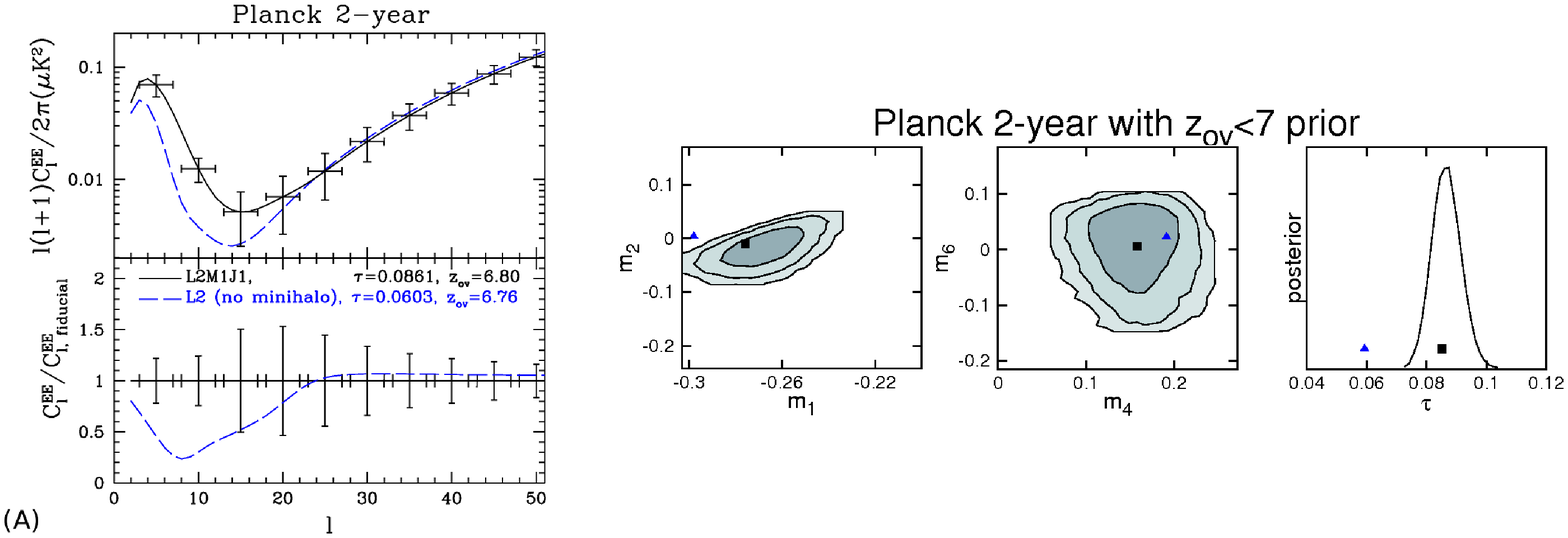}
\includegraphics[width=\textwidth]{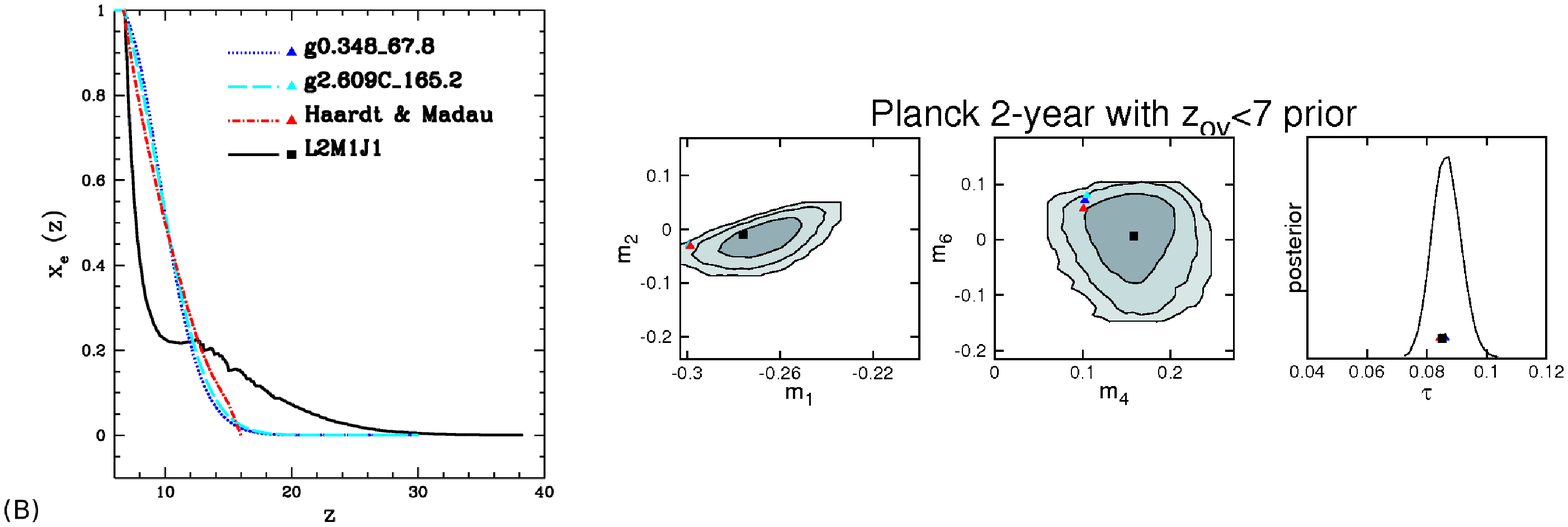}
\caption{
  (A) Detecting the first stars. [{\em left}]: Forecasts of $C_{l}^{\rm EE}$ of cases L2M1J1 (with MHs) 
  and L2 (no MHs) for \emph{Planck}. The error bars are
  estimated \emph{Planck} 2-year $1\sigma$ sensitivity including cosmic
  variance (top panel; \citealt{planck_bluebook}).
  [{\em right}]: Model-selection power of \emph{Planck}. Contours
  represent $1\sigma$ (68\%), $2\sigma$ (95\%) and $3\sigma$ (99.7\%)
  confidence levels from inside out, on marginalized posterior distributions
  of selected parameters ($m_{\mu}$'s and $\tau_{{\rm es}}$)
  using mock data based upon Model L2M1J1 (black square). 
  Case L2 (blue triangle) can be ruled out only
  from the measurement of $\tau_{{\rm es}}$ by \emph{Planck}. The prior
  condition of $z_{{\rm ov}}\le 7$ is applied,
  which rules out early reionization ($z_{{\rm ov}}\gtrsim 8$)
  models.
  (B) Breaking the degeneracy in $z_{\rm ov}$ and
  $\tau_{\rm es}$. [{\em left}]: Ionization histories of various
  models, but with identical $z_{\rm ov}(\simeq 6.8)$ and
  $\tau_{\rm es}(\simeq 0.085)$. The model with MH sources
  (case L2M1J1, black line) stands out
  from almost identical, no-MH models. $g0.348\_67.8$ (no clumping; blue, dotted)
  and $g2.609{\rm C}\_165.2$ (with z-dependent clumping; cyan, dashed) are semi-analytic
  models obtained from equation (A1) of \citet{Iliev2007} with 
  $n=0.1$, and Haardt \& Madau (red, dot-dashed) is from \citet{Haardt2012}.
  [{\em right}]: Hypothesis-testing power of \emph{Planck} on MH-included
  (black square)
  vs. no-MH models (triangles). Contours have the same meaning
  as those in (A). No-MH
  models are clustered and well separated from case L2M1J1 at $\gtrsim
  2\sigma$ confidence level. 
\label{fig:ee}}
\end{figure*}

As we discussed above, current observational constraints 
suggest that reionization was not complete before $z\sim7$.
Imposing this condition as a prior on the
allowed reionization histories $x_m(z)$, we predict that the \emph{Planck} 
mission will clearly detect the era of first stars (Fig.~\ref{fig:ee}). In 
Fig.~\ref{fig:ee} we show a statistical measure of the \emph{Planck} 
sensitivity to detecting the signature of the first stars through the
principal component analysis by
\citet{Hu2003} and \citet[who modified COSMOMC \citep{Lewis2002} to allow generic
  reionization models]{Mortonson2008}\footnote{While we use the
  scheme by
  \citet{Mortonson2008}, we implement the following
  ingredients to optimize the analysis for our
  purpose. First, to apply the late-reionization prior, $z_{\rm 
    ov}\le 7$, we created 7 sets of principal components based on
  $x_{e,\,{\rm fid}}(z) = \frac{40-z}{40-6.5}$ (see Eq.~\ref{eq:xez_pc} for the
  definition of $x_{e,\,{\rm fid}}(z)$), which make $x_{e}(z)$ behave well around
  $z\simeq z_{\rm ov}$. We then use this late-reionization prior to reject any
  sample reionization history with $z_{\rm ov}>7$ when forming the
  Monte-Carlo Markov-chain of varying reionization models. Second, we
  improve the physicality condition, or
  $0\le x_{e}(z)\le 1$ at any z, which was somewhat poorly applied in
  \citet{Mortonson2008}. Whenever a set of $m_\mu$ parameters
  (Eq.~\ref{eq:mmu}) are sampled, we calculate the corresponding $x_{e}(z)$, and when either
  ${\rm min}(x_{e}(z))>−0.04$ or ${\rm max}(x_{e}(z)) <1.04$ is
  violated, we reject that sample. 
  This
  small, 4\% non-physicality in $x_{e}$ is still necessary because of the oscillatory
  nature of $x_{e}(z)$ caused by the limited number of principal
  components, but has only modest effects on the CMB E-mode polarization
  power spectrum $C_{l}^{\rm EE}$.}.
Reionization principal components $\{S_{\mu}(z)\}$ are 
eigen-vectors of the relevant Fisher matrix (evaluated with
an artibrarily chosen fiducial history $x_{e,\,{\rm fid}}(z)$), 
\begin{equation}
F_{ij}\equiv{\displaystyle 
\sum_{l=2}^{l_{{\rm max}}}}\left(l+\frac{1}{2}\right)
\frac{\partial^{2}C_{l}^{{\rm EE}}}{\partial x_{e}(z_{i})\partial
  x_{e}(z_{j})},
\end{equation}
which can be used to describe any generic ionization
history $x_{e}(z)$ with just a small number of modes, such that
\begin{equation}
x_{e}(z)=x_{e,\,{\rm fid}}(z)+{\displaystyle \sum_{\mu=1}^{N_{{\rm
          max}}}}m_{\mu}S_{\mu}(z).
\label{eq:xez_pc}
\end{equation}
The mode amplitude $m_{\mu}$, for a given history $x_{e}(z)$, becomes
\begin{equation}
m_{\mu}=\frac{\int_{z_{\rm min}}^{z_{\rm
        max}} dz \,S_{\mu}(z)\left[x_{e}(z)-x_{e,\,{\rm
        fid}}(z)\right]}{z_{\rm max}-z_{\rm min}}. 
\label{eq:mmu}
\end{equation}
Based on the \emph{Planck} data after 
its full 2 years of planned operation, the narrow posterior distribution 
of allowed $\tau_{{\rm es}}$ values will allow us to distinguish reionization 
models like L2 and L2M1J1 \emph{unambiguously}, and thereby strongly
constrain the available 
reionization models. A high measured measured value of 
$\tau_{\rm  es}>0.085$ will be a clear (if indirect) signature of the first 
stars. 

Finally, we note that the presence of MH sources introduces  
the $x_m$-plateau noted above, which in turn imprints characteristic features in 
$C_{l}^{\rm EE}$. Hence, \emph{Planck} might be able to 
distinguish (albeit at lower statistical significance, of
$\gtrsim 2\sigma$ or $\gtrsim 95\%$) reionization 
models with and without first stars \emph{even} if they have very similar 
values of $\tau_{\rm es}$ and 
$z_{\rm ov}$ (Fig.~\ref{fig:ee}B). Full reionization simulations like
ours find it hard to satisfy
both of these observational constraints without including a significant contribution 
from the first stars, but some semi-analytical models
(\citealt{Haiman2006,Haardt2012}; $g0.348\_67.8$ and $g2.609{\rm
  C}\_165.2$ in Fig.~\ref{fig:ee}B which are of unnaturally large
gaps in relative efficiencies of LMACH and HMACH)
do find such scenarios. However, all such models lack the plateau feature 
in $x_m(z)$, regardless of the details of the assumed physics, and
reside in a narrow 
window of ${m_\mu}$-parameter space adjacent to that occupied by our
no-MH cases, as demonstrated in Fig.~\ref{fig:ee}B.

In summary, \emph{Planck} is capable of distinguishing with high confidence
between definitive classes of reionization scenarios allowed by the
current constraints, and thereby significantly restricting the available
parameter space. \emph{Planck} will either probe the signature of the
first stars, or show that the first stars had a negligible impact on
reionizaion. Once these first results confirm the role of the first 
stars, simulations of the type presented here can be used  
to study other observable quantities and thus deepen our understanding
of the early universe.

\acknowledgments
K.A. was supported in part by NRF grant funded by the Korean government
MEST (No. 2009-0068141, 2009-0076868, 2012R1A1A1014646,
2012M4A2026720). ITI was supported by The Southeast 
Physics Network (SEPNet) and the Science and Technology Facilities
Council grants ST/F002858/1 and ST/I000976/1. This study was supported
in part by the Swedish Research Council grant 2009-4088, U.S. NSF grants
AST-0708176 and AST-1009799, NASA grants NNX07AH09G, NNG04G177G and
NNX11AE09G, and Chandra grant SAO TM8-9009X. The authors acknowledge
the TeraGrid and the Texas Advanced Computing Center (TACC) at The
University of Texas at Austin (URL: http://www.tacc.utexas.edu), and
the Swedish National Infrastructure for Computing (SNIC) resources
at HPC2N (Ume\aa, Sweden) for providing HPC and visualization resources
that have contributed to the results reported within this
paper. We acknowledge A. Lewis, A. Liddle and M. Mortonson
for scientific and technical input on \emph{Planck} forecasts (Lewis, Liddle)
and COSMOMC modified for reionization principal components (Mortonson).

\newpage

%%table

\begin{table*}[ht]
{\linespread{1.}
\begin{center}
\caption{Reionization simulation source halo properties and global history results.
$M_{{\rm III},\,*}$ and $J_{{\rm LW,\, th}}$ are in units of solar
mass ($M_\odot$) and $10^{-21}\,{\rm erg}\,{\rm s}^{-1}\,{\rm cm}^{-2}\,{\rm sr}^{-1}$, respectively. 
\emph{Note}: MH efficiencies $g_{\gamma,\,{\rm MH}}$
($f_{\gamma,\,{\rm MH}}$) quoted here are for the minimum-mass halo
assumed to contribute, $10^5\,M_\odot$, which is roughly comparable to
the average value for the minihalos integrated over the halo mass function. The efficiency of the MH of mass $M$ is
obtained simply by multiplying $\left( \frac{10^{5} \,M_{\odot}}{M} \right)$ to the
quoted $g_{\gamma,\,{\rm MH}}$ ($f_{\gamma,\,{\rm MH}}$).
\label{table:param}}
%%\begin{tabular}{crrrrrrrr}
\begin{tabular}{|>{\footnotesize}p{3.4cm}|>{\footnotesize}p{0.8cm}|>{\footnotesize}p{0.8cm}|>{\footnotesize}p{1cm}|>{\footnotesize}l|>{\footnotesize}l|>{\footnotesize}l|>{\footnotesize}l|>{\footnotesize}p{3.5cm}|}
\hline
case & $g_{\gamma,\,{\rm H}}$  $(f_{\gamma,\,{\rm H}})$ & $g_{\gamma,\,{\rm L}}$  $(f_{\gamma,\,{\rm L}})$ & $g_{\gamma,\,{\rm MH}}$  $(f_{\gamma,\,{\rm MH}})$ & $M_{{\rm III},\,*}$ & $J_{{\rm LW,\, th}}$ & $z_{{\rm ov}}$ & $\tau_{{\rm es}}$ & $m_{1}, m_{2}, ..., m_{7}$\tabularnewline
\hline
$g8.7\_130{\rm S}$ \newline (L1) & 8.7 \newline (10) & 130 \newline (150) &  $\cdot$ & $\cdot$ & $\cdot$ & 8.40 & 0.0841 & -0.298, -0.0267, 0.289, \newline 0.115, 0.0975, 0.0918, \newline -0.0548\tabularnewline
\hline 
$g8.7\_130{\rm S}\_M300\_J0.05$ (L1M1J2) & 8.7 \newline (10) & 130 \newline (150) & 5063 \newline (1013) & 300 & 0.05 & 8.41 & 0.0934 & -0.283, -0.0222, 0.268, \newline 0.121, 0.0828, 0.0897, \newline -0.0565\tabularnewline
\hline 
$g8.7\_130{\rm S}\_M100\_J0.05$ (L1M2J2) & 8.7 \newline (10) & 130 \newline (150) & 1687.7 \newline (337.7) & 100 & 0.05 & 8.41 & 0.0910 & -0.288, -0.0234, 0.274, \newline 0.120, 0.0868, 0.0908, \newline -0.0558\tabularnewline
\hline 
$g8.7\_130{\rm S}\_M300\_J0.01$ (L1M1J3) & 8.7 \newline (10) & 130 \newline (150) & 5063 \newline (1013) & 300 & 0.01 & 8.41 & 0.0874 & -0.293, -0.0236, 0.283, \newline 0.118, 0.0952, 0.0919, \newline -0.0541\tabularnewline
\hline 
$g8.7\_130{\rm S}\_M100\_J0.01$ (L1M2J3) & 8.7 \newline (10) & 130 \newline (150) & 1687.7 \newline (337.7) & 100 & 0.01 & 8.41 & 0.0861 & -0.295, -0.0247, 0.285, \newline 0.117, 0.0962, 0.0918, \newline -0.0545\tabularnewline
\hline 
$g1.7\_8.7{\rm S}$ \newline (L2) & 1.7 \newline (2) & 8.7 \newline (10)  & $\cdot$ & $\cdot$ & $\cdot$ & 6.76 & 0.0603 & -0.298, 0.00402, 0.372, \newline 0.191, 0.0446, 0.0229, \newline -0.0416\tabularnewline
\hline 
$g1.7\_8.7{\rm S}\_M300\_J0.1$ (L2M1J1) & 1.7 \newline (2) & 8.7 \newline (10) & 5063 \newline (1013) & 300 & 0.1 & 6.80 & 0.0861 & -0.276, -0.00969, 0.302, \newline 0.158, 0.0260, 0.00619, \newline -0.0349\tabularnewline
\hline 
$g1.7\_8.7{\rm S}\_M300\_J0.05$ (L2M1J2) & 1.7 \newline (2) & 8.7 \newline (10) & 5063 \newline (1013) & 300 & 0.05 & 6.79 & 0.0788 & -0.281, -0.00479, 0.323, \newline 0.170, 0.0285, 0.00893, \newline -0.0377\tabularnewline
\hline
\end{tabular}
\end{center}
} %% linespread ended
\end{table*}
%% table ended

%\bibliographystyle{apj}
%\bibliography{refs_2012apr}

\end{document}